\documentclass[aps,pra,amssymb,amsfonts,floatfix]{revtex4}

\usepackage{graphicx}% Include figure files
\usepackage{dcolumn}% Align table columns on decimal point
\usepackage{bm}% bold math
\usepackage{mathbbol}

\begin{document}
\title{Quantum interference-induced stability of repulsively bound pairs of excitations}

\author{Lea F. Santos$^{1}$ and M. I. Dykman$^{2}$} 
\affiliation{$^1$~Department of Physics, Yeshiva University, New York, NY, 10016, USA\\
$^2$~Department of Physics and Astronomy, Michigan State University, East Lansing, MI 48824, USA}
\email{lsantos2@yu.edu}

\date{\today}

\begin{abstract}
We study the dynamics of two types of pairs of excitations which are bound despite their strong repulsive interaction. One corresponds to doubly occupied sites in one-dimensional Bose-Hubbard systems, the so-called doublons. The other is pairs of neighboring excited spins in anisotropic Heisenberg spin-1/2 chains. We investigate the possibility of decay of the bound pairs due to resonant scattering by a defect or due to collisions of the pairs. We find that the amplitudes of the corresponding transitions are very small. This is a result of destructive quantum interference and explains the stability of the bound pairs.
\end{abstract}

\date{\today}
%\pacs{...}
\maketitle

\section{Introduction}

Cold atoms trapped in optical lattices offer high degree of control over such factors as the interaction strength, level of disorder, and system geometry. This makes it possible to simulate, in a controlled fashion, various models of condensed matter physics~\cite{Greiner2008,Bloch2008}.  Bose~\cite{Greiner2002,Winkler2006,Trotzky2012} and Fermi~\cite{Schneider2008,Jordens2008,Strohmaier2010} Hubbard models are among the studied ones, and spin-1/2 systems are receiving increasing attention~\cite{Trotzky2008,Simon2011,Chen2011}. Another significant advantage of cold atom systems is their weak coupling to the environment. This makes it possible to study their nearly dissipation-free evolution for long times, which in turn has led to the renewed interest in the far-from-equilibrium dynamics of many-body quantum systems.

Several works have been recently dedicated to the formation~\cite{Winkler2006,Schecter2012,TokunoARXIV}, dynamics~\cite{Nygaard2008,Hassler2010,Dias2010,Hofmann2012,DeuchertARXIV}, and relaxation~\cite{Strohmaier2010,Sensarma2010,Chudnovskiy2012} of bound pairs (BPs) of atoms that can exist even in the presence of repulsive interactions. They correspond to doubly occupied lattice sites and have been called doublons. Doublons have been observed in Bose~\cite{Winkler2006} and Fermi~\cite{Strohmaier2010,Sensarma2010} Hubbard models with short range interaction. When the interaction is strong, be it attractive or repulsive, an energy gap can emerge between the band of doublons and the band where each atom is on a singly-occupied lattice sites. This gap, together with the isolation of the system, results in the long lifetime of the BPs in ideal systems. In some sense, this lifetime is even unexpectedly long. Similarly, BPs of parallel spins can emerge in spin-1/2 systems with strongly anisotropic  nearest-neighbor interaction, commonly referred to as XXZ models. In a spin chain where all spins point down due to a strong magnetic field, for example, two up-spins on neighboring sites form a long-lived BP. 

All realistic systems have defects. They can lead to scattering of BPs. Understanding the stability of BPs in the presence of such scattering is therefore of central interest. For BPs in spin-1/2 chains with a single on-site defect and large anisotropic exchange interaction, this problem was addressed in our previous articles \cite{SantosDykman2003,DykmanSantos2003}. In this work we extend the analysis to doublons in one-dimensional Bose-Hubbard (BH) system and provide a comparison with the previous results. We also discuss the stability of BPs against many-particle scattering in ideal atomic and spin chains. The presence of an atom on a lattice site in the BH model or of a flipped spin in the XXZ chain is generally referred to as an excitation. 

We consider one-dimensional systems where the energy of the on-site positive Hubbard interaction or the nearest-neighbor $S^z-S^z$ spin interaction largely exceeds the bandwidth of single-particle excitations. In the BH model, the bandwidth is determined by the hopping integral, whereas in the XXZ model it is determined by the in-plane spin coupling. This condition guarantees the energy separation of the BP band from the band of unbound two-particle excitations. We then analyze the case where the system has a site defect with an excess energy that also largely exceeds the bandwidth of single-particle excitations. Such a defect would be expected to cause strong scattering of BPs if a BP could resonantly decay into an excitation localized on the defect and a propagating excitation. This resonance can be achieved by tuning the defect site energy to a value close to the interaction energy. We show, however, that such inelastic scattering is largely suppressed. This is a consequence of destructive quantum interference.

We also discuss the decay of BPs in a defect-free chain. The goal is to understand what keeps the BPs stable when they collide with each other~\cite{Winkler2006}. In the anisotropic Heisenberg model, energy conservation 
allows two colliding bound spin pairs to decay into a bound three-spin state and an unbound spin excitation. We show, however, that such scattering is also strongly suppressed by destructive quantum interference. For the BH model, energy conservation requires a multi-particle collision for the emergence of a bound triple of excitations. 

This paper is organized as follows. The two models analyzed are described in Sec.~\ref{SecModels}. We consider strong interaction and a small number of excitations. The analytical studies developed here are based on Bethe ansatz~\cite{Bethe1931,Alcaraz1987,Karbach1997} and perturbation theory, as explained in Secs.~\ref{SecOne} and \ref{SecTwo}. The core of the work appears in Secs.~\ref{SecBPdef} and \ref{TwoBPs}. In Sec.~\ref{SecBPdef}, we show that a BP has a parametrically small amplitude to split by sending an excitation to the defect site, and in Sec.~\ref{TwoBPs}, we show that a two-BP state essentially does not couple with a bound three-excitation state, even though both scattering processes are allowed by energy conservation. The BPs are almost elastically reflected from a defect site and from each other. 

\section{Models}
\label{SecModels}

We study one-dimensional systems with $L$ sites and periodic boundary conditions (closed chains) described by the BH and XXZ models. In both cases we assume strong interaction and take $\hbar =1$.

\subsection{Bose-Hubbard model}

In the BH model the interaction occurs on the same site and the Hamiltonian is given by,
\begin{equation}
H^{\text{BH}} =  \sum_{j=1}^{L} \varepsilon_j n_j  
+ \frac{U}{2} \sum_{j=1}^{L} n_{j} \left(n_{j} -1 \right)
- J \sum_{j=1}^{L} \left(b_j^{\dagger} b_{j+1} + b_{j+1}^{\dagger} b_{j} \right).
\label{HamBH}
\end{equation}
Here, $b_j$ and $b_j^{\dagger}$ are respectively bosonic annihilation and creation operators on site $j$, $n_j= b_j^{\dagger} b_j$ is the corresponding number operator, $J$ is the nearest-neighbor hopping integral, and $U$ is the energy of the on-site particle interaction. The parameter $\varepsilon_j$ is a site-dependent energy accounting for inhomogeneous external potentials, which can be experimentally controlled~\cite{Fallani2007}. We assume that this background potential is the same for all sites, except for $j=d$, referred to as the defect site, so that $\varepsilon_j=\varepsilon + g  \delta_{j,d}$, with $\varepsilon >0$.
The BH Hamiltonian conserves the total number of atoms (excitations) $N$, the dimension of each subspace being $(L-1+N)!/[(L-1)! N!]$.

\subsection{XXZ model}

The anisotropic spin-1/2 Heisenberg model, also known as the XXZ model, is described by the Hamiltonian,
\begin{equation}
H^{\text{XXZ}} =  \sum_{j=1}^{L} B_j S_j^z  
+ J_z \sum_{j=1}^{L} S_j^z S_{j+1}^z
- J_{xy} \sum_{j=1}^{L} \left(S_j^x S_{j+1}^x + S_j^y S_{j+1}^y \right).
\label{HamXXZ}
\end{equation}
Here, $S^{x,y,z}_j = \sigma^{x,y,z}_j/2$ are the spin operators on site $j$, $\sigma^{x,y,z}_j$ being the Pauli matrices. The parameter $B_j = B + g \delta_{j,d}$, with $B >0$, gives the Zeeman splitting of each site, as determined by a static magnetic field in the $z$ direction; it is the same for all sites, except for the defect site $d$. 
The Hamiltonian contains only nearest-neighbor exchange with coupling constants $J_z$ and $J_{xy}$; $S_j^z S_{j+1}^z$ is the Ising-type interaction and $S_j^x S_{j+1}^x + S_j^y S_{j+1}^y$ is the flip-flop term~\cite{footnote}.
We assume a large effective magnetic field, $B\gg |J_z|$ pointing up in the $z$ direction, so that the ground state has all spins pointing down, independently of the sign of $J_z$. A spin pointing up corresponds then to an excitation.

The Heisenberg Hamiltonian conserves the total number of spins in the $z$ direction, ${\cal S}^z = \sum_{j=1}^L S_j^z$, and therefore it consists of blocks of dimension  $L!/[(L-N)! N!]$, where $N$ is the number of up-spins. The defect-free system is integrable and can be solved with the Bethe ansatz~\cite{Bethe1931,Alcaraz1987,Karbach1997}. It can also be mapped onto a system of spinless fermions~\cite{Jordan1928} or hardcore bosons~\cite{Holstein1940}. In the presence of a single defect with excess energy $g \sim J_{xy},J_z$, the system becomes chaotic, as attested by the Wigner-Dyson form of the level spacing distribution~\cite{Santos2004,Barisic2009,Gubin2012}. However, if the defect is very large, the system behaves as a chain with open boundary conditions, where the excitations move freely bouncing back and forth from the defect site. In this situation, analytical studies are again viable.

In the next section, by combining perturbation theory and Bethe ansatz, we derive analytically the eigenvalues and eigenstates for the trivial subspace of a single excitation, where the BH and XXZ models are identical. It serves as an introduction for the sections where two excitations are considered.
The basis vectors used are states where each excitation is on a single site. They are eigenstates of the interaction part of Hamiltonians (\ref{HamBH}) and (\ref{HamXXZ}), as for example, $|0010\rangle$ for $H$ (\ref{HamBH}) or equivalently $|\downarrow \downarrow \uparrow \downarrow \rangle $ for $H$ (\ref{HamXXZ}).
The table below gives their energies ${\cal E}_N$  for the cases where the excitations are far from each other and away from the defect site.
\begin{table}[h]
\begin{tabular}{lr}
\\ [-6pt]
 BH model & XXZ model \\ [2pt]
\hline \\ [-6pt]
  ${\cal E}_0^{\text{BH}} = 0 $ & \hspace{1 cm}${\cal E}_0^{\text{XXZ}} = -BL/2 -g/2+J_z (L-1)/4$ \\ [5pt]
 \hline  \\ [-6pt]
  ${\cal E}_N^{\text{BH}} = N \varepsilon $ & \hspace{1 cm}${\cal E}_N^{\text{XXZ}} = {\cal E}_0^{\text{XXZ}} + N B - N J_z$ \\ [5pt]  
   \hline
\end{tabular}
\caption{Energies of the basis vectors for the BH and XXZ models, which correspond to $N$ localized excitations far from each other and away from the defect site.}
\label{tableEnergies}
\end{table}

\section{One excitation and one defect}
\label{SecOne}

We start with a single excitation in a chain with a large-energy defect, $|g|>>|J|,|J_{xy}|$. The defect breaks translational symmetry. The scenario becomes similar to that of a system with open boundaries, the excitation either moves freely along the chain, being reflected when it gets next to $d$, or it is localized on the defect site. The results shown below are independent of the signs of $g$ and $J,J_{xy}$, so we assume them positive~\cite{SantosDykman2003,DykmanSantos2003}.

\subsection{A single freely propagating excitation}

For an excitation away from the defect, we look for approximate solutions of the type,
\begin{equation}
\psi_{1,\text{free}}^{\text{BH}} = \sum_{x=d+1}^{d-1+L} a(x) \phi(x), \qquad a(d)=a(d+L)=0,
\label{psi1}
\end{equation}
where $\phi(x)$ indicates the basis vector where the excitation is on site $x$ and $a(x)$ is the  probability amplitude of this state. 

The effect of the BH Hamiltonian (\ref{HamBH}) on $\phi(x)$ is
\begin{equation}
H^{\text{BH}} \phi(x) = {\cal E}_1^{\text{BH}} \phi(x) - J [\phi(x-1) + \phi(x+1)] , 
\label{phi1}
\end{equation}
where ${\cal E}_1^{\text{BH}} = \varepsilon$ (cf. Table~\ref{tableEnergies}).
Substituting Eqs.~(\ref{psi1}) and (\ref{phi1}) into the Schr\"odinger equation $H \psi_1 = E_1 \psi_1$ and selecting the terms proportional to $\phi(x)$, we obtain
\begin{equation}
{\cal E}_{1}^{\text{BH}} a(x) - J [a(x-1) + a(x+1)] = E_{1,\text{free}}^{\text{BH}} a(x).
\end{equation}

We now make an ansatz for $a(x)$, taking into account the reflection of the excitation by the defect site,  so that
\[
a(x) = C e^{i \theta x} + C' e^{-i\theta x}, 
\]
with the condition $a(d) \equiv a(d+L)=0$.
This leads to the energies,
\begin{equation}
E_{1,\text{free}}^{\text{BH}}(\theta)={\cal E}_1^{\text{BH}} -2J \cos \theta , 
\end{equation}
and the eigenstates,
\begin{equation}
\psi_{1,\text{free}}^{\text{BH}} (\theta) = \sum_{x=d+1}^{d-1+L} A \sin[\theta (x-d)] \phi(x),
\end{equation}
where $A$ is a normalization factor and
\[
\theta = \frac{\pi k}{L} \hspace{0.4 cm} \text{with} \hspace{0.4 cm} k=1,2, \ldots (L-1).
\]

The same eigenstates are obtained for the XXZ  model, the eigenvalues being
\begin{equation}
E_{1,\text{free}}^{\text{XXZ}}(\theta) ={\cal E}_1^{\text{XXZ}} -J_{xy} \cos \theta,
\end{equation}
with ${\cal E}_1^{\text{XXZ}}$ given in Table~\ref{tableEnergies}.

\subsection{An excitation localized on the defect}

Since $g$ is very large, the localization length of the excitation on the defect site is very small, which leads to the trivial state, $\psi_{1,\text{loc}}^{\text{BH}} = \phi(d)$, with energy 
\begin{equation}
E_{1,\text{loc}}^{\text{BH}} = {\cal E}_1^{\text{BH}} + g + \frac{2 J^2}{g} .
\end{equation}
The last term above is a second-order perturbation theory correction, which when taken into account leads to excellent agreement with the numerics. It is derived from a virtual transition of the excitation on the defect to the site next to it and back to the defect site. For the XXZ  model the energy of the localized state is given by
\begin{equation}
E_{1,\text{loc}}^{\text{XXZ}}={\cal E}_1^{\text{XXZ}} +g + \frac{J_{xy}^2}{2g}.
\end{equation}

\section{Two excitations and one defect}
\label{SecTwo}

In the case of two excitations, we look for solutions of the type

\begin{equation}
\psi_2^{\text{BH}} = \sum_{x\leq y} a(x,y) \phi(x,y), \hspace{2.0 cm} 
\psi_2^{\text{XXZ}} = \sum_{x< y} a(x,y) \phi(x,y),
\label{psi2}
\end{equation}
where $\phi(x,y)$ indicates the state where the excitations are on site $x$ and $y$ and $a(x,y)$ is its probability amplitude. In the BH model, there are states with $x=y$ where two excitations are on the same site, while in the XXZ  model, the excitations can reach neighboring sites, but cannot be on the same site, so $x\neq y$.

We are interested in the case where the interaction is strong, $|U| \gg J$ for the BH model and $|J_z| \gg J_{xy}$ for the XXZ  model, which creates two well separated bands of energy. In one band, the excitations are noninteracting and move freely along the chain; in the other, the excitations are bound together due to the interaction and move slower as a single heavier particle~\cite{Winkler2006,SantosDykman2003,DykmanSantos2003}. The cause of the onset of such pair of bound excitations is the difference in energy between the two bands. States with freely propagating excitations in the BH-model have energy $\sim {\cal E}_2^{\text{BH}}$, whereas the BP-states have energy $\sim {\cal E}_2^{\text{BH}} + U$. For the XXZ model, the energies of these states are respectively $\sim {\cal E}_2^{\text{XXZ}}$ and $\sim {\cal E}_2^{\text{XXZ}} + J_z$. BPs can therefore emerge for either attractive or repulsive interactions. From now on, we assume repulsive interaction, $U,J_z>0$. 

\begin{figure}[htb]
\vskip 0.7cm
\includegraphics[width=3.5in]{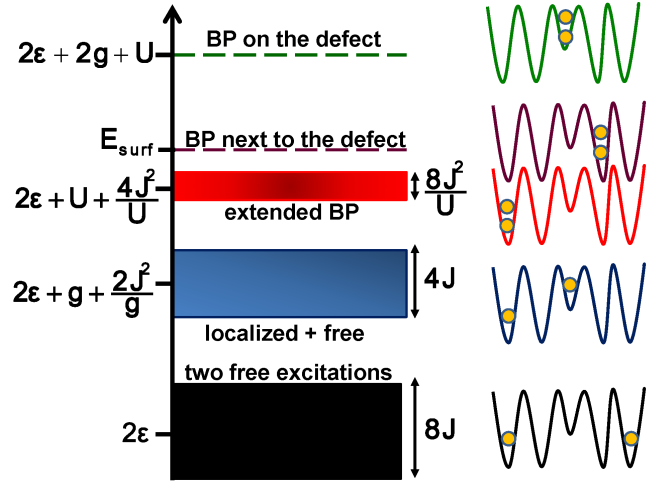}
\caption{Two-excitation energy spectrum for the BH model when $U\gg J$ and $U/2 < g < U$. We assume $g,J,U>0$. The band centered at ${\cal E}_2^{\text{BH}}=2\varepsilon$ is formed by noninteracting excitations. The band centered at $2\varepsilon +g+2J^2/g$ is formed by states with one excitation localized on the defect site and the other moving freely. The narrow band centered at $2\varepsilon + U +4J^2/U$ is formed by propagating
pairs of bound excitations. The dashed line above the BP-band indicates the localized states where the BPs are next to the defect; $E_{\rm surf} = 2\varepsilon +U+2J^2U/[g(U-g)]$ (see text). The dashed line at  highest energy indicates the BP on the defect site.}
\label{fig:diagram}
\end{figure}

In addition to strong interaction, we consider also a defect with a large site energy, so that $g \gg J, J_{xy}$ and $U-g\gg J$ for the BH model or equivalently $J_z-g\gg J_{xy}$ for the XXZ model. This leads to a third energy band placed between the previous two. Here, one excitation is trapped on the defect site and the other moves freely along the chain. The energy of this band is $\sim {\cal E}_2 + g$. Thus, the three bands are  separated in energy, as illustrated in Fig.~\ref{fig:diagram}. The dashed lines in the diagram indicate two kinds of localized states that emerge close to the defect. These states and the energy bands are described in the next subsections.

\subsection{Two freely propagating excitations} 

The lowest energy band in Fig.~\ref{fig:diagram} is composed of states with noninteracting excitations; the excitations are sufficiently far apart from each other and from the defect site to move freely.  
In this case, the Schr\"odinger equation for the BH Hamiltonian leads to 
\begin{equation}
{\cal E}_2^{\text{BH}} a(x,y) - J [a(x-1,y) + a(x,y-1) + a(x+1,y)+a(x,y+1)] = E_{2,\text{free}}^{\text{BH}} a(x,y),
\label{S2}
\end{equation}
where ${\cal E}_2^{\text{BH}}=2\varepsilon$ (cf. Table~\ref{tableEnergies}). The defect works as a barrier that reflects the excitations before they reach site $d$. Due to energy conservation, the excitations are also reflected before reaching the same site. An appropriate ansatz for the probability amplitude away from the defect is then
\begin{equation}
a(x,y) = C e^{i \theta x} e^{- i \theta' y} + C' e^{-i\theta x} e^{i \theta' y},
\label{axy}
\end{equation}
with the condition $a(d,y)=a(x,d+L)=a(x,x)=0$. By substituting (\ref{axy}) back into (\ref{S2}), we obtain the eigenvalues
\begin{equation}
E_{2,\text{free}}^{\text{BH}}(\theta,\theta') = {\cal E}_2^{\text{BH}} - 2 J (\cos \theta + \cos \theta'),
\end{equation}
constituting a band of width $8J$.

There are $(L-1)(L-2)/2$ states with two freely propagating excitations for the BH model and $(L-2)(L-3)/2$ states for the Heisenberg model. For the latter, the energies are
\begin{equation}
E_{2,\text{free}}^{\text{XXZ}}(\theta,\theta') = {\cal E}_2^{\text{XXZ}} - J_{xy} (\cos \theta + \cos \theta').
\end{equation}

\subsection{An excitation localized on the defect and a freely propagating excitation}
\label{SubSecLDP}

The energy of the states of the BH model where one excitation is localized on the defect and the other moves freely between sites $d+1$ and $d+L-1$ (referred to here as loc-free-states) differ by $g$ from the states where both excitations propagate freely and by $U-g$ from BPs. By taking into account the energy of the excitation on the defect up to second order of perturbation theory, the eigenvalues and eigenvectors of the loc-free-band can be found from the equation
\begin{equation}
\label{eq:loc_free}
\left[ {\cal E}_2^{\text{BH}} +g + \frac{J^2}{g}\left(2-\delta_{y,d\pm 1}\frac{U^2-g^2+4Ug}{U^2-g^2}\right) \right] a(d,y) - J [a(d,y-1) + a(d,y+1)] = E_{2,\text{loc-free}}^{\text{BH}} a(d,y).
\end{equation}
To the lowest order in $J^2/g, J^2U/(U^2-g^2)$, we seek the solution in the form
\begin{equation}
a(d,y) = C e^{i\theta y} + C' e^{-i\theta y},
\label{ldp}
\end{equation}
with the condition $a(d,d)=a(d,d+L)=0$. The scenario is very similar to that of a free excitation described in Sec.~\ref{SecOne}. The eigenvalues are well approximated by
\begin{equation}
E_{2,\text{loc-free}}^{\text{BH}}(\theta)= \left[ {\cal E}_2^{\text{BH}}+ g + \frac{2J^2}{g} \right] - 2 J \cos \theta,
\end{equation}
and the eigenstates have the form
\begin{eqnarray}
&& \psi_{2,\text{loc-free}}^{\text{BH}}(\theta) = \sum_{y=d+1}^{d-1+L} A \sin[\theta (y-d)] \phi(d,y), 
\label{Psilocfree}\\
&& \theta = \frac{\pi k}{L} \hspace{0.4 cm} \text{with} \hspace{0.4 cm} k=1,2, \ldots (L-1),
\label{thetalocfree}
\end{eqnarray}
where $A$ is a normalization factor.

The results for the Heisenberg model are similar [see Table~\ref{tableTwoExcs} for a mapping between the two models], except that here there are only $L-3$ states, since the boundary conditions are $a(d,d+1)=a(d,d-1+L)=0$.

\subsection{Extended and localized bound pairs}

In the leading order of perturbation theory, transport of a pair of bound excitations through the chain occurs  via an intermediate virtual transition where the pair splits and then recombines again. The BP behaves as a single heavy free excitation in an open chain. In the BH model, it moves from site $d+1$ to $d+L-1$ with effective hopping integral $2J^2/U$. The eigenvalues and eigenvectors can be found from the equations
\begin{eqnarray}
&&\left[ {\cal E}_2^{\text{BH}} + U + \frac{4J^2}{U} \right] a(x,x) + \frac{2J^2}{U} [a(x-1,x-1) + a(x+1,x+1)] = E_{2,\text{BP}}^{\text{BH}} a(x,x), 
\label{BPextended} \\
&&\left[ {\cal E}_2^{\text{BH}} \!+\! U \!+\! \frac{4J^2}{U} \!+\! \Delta \right] a(d+1,d+1) + \frac{2J^2}{U} a(d+2,d+2)  = E_{2,\text{BP}}^{\text{BH}} a(d+1,d+1), 
\label{BPplus} \\
&&\left[ {\cal E}_2^{\text{BH}} \!+\! U \!+\! \frac{4J^2}{U} \!+\! \Delta \right] a(d\!+\!L\!-\!1,d\!+\!L\!-\!1) \!+\! \frac{2J^2}{U} a(d\!+\!L\!-\!2,d\!+\!L\!-\!2)  = E_{2,\text{BP}}^{\text{BH}} a(d\!+\!L\!-\!1,d\!+\!L\!-\!1), 
\label{BPminus}
\end{eqnarray}
where
\[
\Delta=\frac{2J^2}{U-g} - \frac{2J^2}{U} = \frac{2J^2 g}{U(U-g)}
\]
is the energy difference between the states right next to the defect, $\phi(d+1,d+1)$ and $\phi(d+L-1,d+L-1)$, and the other BPs. 
This difference arises because, in the second order of perturbation theory, the virtual dissociation of a BP with one excitation hopping onto a defect site gives
a different energy denominator than in the case where the virtual transition is made onto a regular site. 

The ansatz
\[
a(x,x) = C e^{i\theta x} + C' e^{-i\theta x}, \qquad d<x<d+L,
\]
in Eq.~(\ref{BPextended}) leads to the eigenvalues
\begin{equation}
E_{2,\text{BP}}^{\text{BH}}(\theta) = \left[ {\cal E}_2^{\text{BH}} + U + \frac{4J^2}{U} \right] + \frac{4J^2}{U} \cos \theta.
\end{equation}
They form a narrow energy band of width $8J^2/U$.

Equation~(\ref{BPextended}) has to be compatible with Eq.~(\ref{BPplus}) for $x=d+1$ and with Eq.~(\ref{BPminus}) for $x=L+d-1$. This leads to a system of two linear equations for the coefficients $C$ and $C'$. The equations have a nontrivial solution for 
$\theta$ satisfying the condition
\begin{equation}
\label{eq:theta_BPs}
f(\theta) = f(-\theta), \hspace{0.6 cm} f(\theta)=
e^{i \theta (L-2)} \left[\Delta  - \frac{2J^2}{U} e^{i\theta} \right]^2.
\end{equation}

As the defect energy increases, two states eventually split off the band of extended BP-states. They correspond to surface states, where the excitations are localized primarily next to the defect, 
on sites $d+1$ and $d+L-1$.
In the thermodynamic limit, $L \rightarrow \infty$, the energies of these states, $E_{\text{surf}}$, are derived from the two imaginary roots of Eq.~(\ref{eq:theta_BPs}), 
\[
\text{Im}~ \theta = \pm \ln \left( \frac{g}{U-g} \right),
\]
which appear for $U> g>U/2$. These energies are shown in Fig.~\ref{fig:diagram} with a dashed line.
We then have two surface states and $L-3$ extended BPs where the pairs move between
sites $d+2$ and $d-2+L$. 
A similar scenario emerges for the XXZ model. When $J_z> g>J_z/2$ there are two states with excitations localized next to the defect, on sites $(d+1,d+2)$ and $(d+L-2,d+L-1)$, and $L-4$ extended BP eigenstates. 

A localized state of higher energy appears when a BP is placed on the defect site. For the BH model, this is a single state with energy ${\cal E}_2^{\text{BH}} + U + 2g$. For the Heisenberg system there are two such states with energy ${\cal E}_2^{\text{XXZ}} + J_z + g$; they have excitations on sites $(d,d+1)$ and on $(d,d+L-1)$. These states can form symmetric and antisymmetric linear superpositions \cite{SantosDykman2003,Santos2003}.

Table~\ref{tableTwoExcs} summarizes the results for both models for the loc-free-band and the band of extended BPs. These states together with the BPs localized next to the defect are the main parties to the quantum interference effects analyzed in Sec.~\ref{SecBPdef}. 
\begin{table}[h]
\begin{tabular}{lr}
\\ [-6pt]
 BH model & XXZ model \\ [2pt]
 \hline \\ [-6pt]
  $ U \gg J $ and $U/2 <g < U$ & $ J_z \gg J_{xy} $ and $J_z/2 <g < J_z$ \\ [5pt]
\hline \\ [-6pt]
\multicolumn{2}{c}{One excitation localized on the defect and one freely propagating excitation} \\ [2pt]
\hline \\ [-6pt]
  $ E_{2,\text{loc-free}}^{\text{BH}}= {\cal E}_2^{\text{BH}}+ g + 2J^2/g  - 2 J \cos \theta $ & \hspace{1 cm} $E_{2,\text{loc-free}}^{\text{XXZ}}={\cal E}_2^{\text{XXZ}} +g + J_{xy}^2/(2 g) -J_{xy} \cos \theta$ \\ [5pt]
\hline \\ [-6pt]
\multicolumn{2}{c}{Extended bound-pairs of excitations} \\ [2pt]
\hline \\ [-6pt]
  $ E_{2,\text{BP}}^{\text{BH}}=  {\cal E}_2^{\text{BH}} + U + 4J^2/U  + (4J^2/U) \cos \theta $ & \hspace{0.6 cm} $E_{2,\text{BP}}^{\text{XXZ}}=  {\cal E}_2^{\text{XXZ}} + J_z + J_{xy}^2/(2J_z)  + [J_{xy}^2/(2J_z)] \cos \theta $ \\ [5pt]     
   \hline
\end{tabular}
\caption{Energy spectrum of loc-free and extended BP-states in a chain with a defect and two excitations. The energies of basis vectors with two separated excitations away from the defect for the Bose-Hubbard and the anisotropic Heisenberg models, ${\cal E}_2^{\text{BH}}$ and ${\cal E}_2^{\text{XXZ}}$, respectively, are given in Table~\ref{tableEnergies}. }
\label{tableTwoExcs}
\end{table} 

\section{Antiresonance stabilization of bound pairs in a chain with a defect}
\label{SecBPdef}

In physical systems of interest for the studies of doublons, the defect energy can often be tuned. These systems display a somewhat counterintuitive behavior if the energy of a loc-free state resonates with the energy of a BP away from the defect, that is if $g$ is close to $U$ in the BH-model. In this case the band of extended BPs lies inside the (much broader) band of loc-free states. One could then expect hybridization of the two types of states. In other words, a BP could approach the defect and resonantly scatter into an excitation localized on the defect and a propagating single excitation, or vice versa, a propagating single excitation could resonantly scatter off the excitation on the defect, and then the two excitations would move away from the defect as a bound pair. Both scenarios are allowed by energy conservation, but they have very small probabilities to occur, because of a destructive quantum interference.

When $g\approx U\gg J$, BPs localized immediately next to the defect are directly coupled to the loc-free-states. They form a new hybrid band denoted below with the subscript ``hyb''. The eigenstates and eigenvalues are obtained from the equations,
\begin{eqnarray}
&&\left[ {\cal E}_2^{\text{BH}} + g  \right] a(d,y) - J [a(d,y-1) + a(d,y+1)] 
= E_{2,\text{hyb}}^{\text{BH}} a(d,y), 
\label{HYBextended} \\
&&\left[ {\cal E}_2^{\text{BH}} + U  \right] a(d+1,d+1) - \sqrt{2} J a(d,d+1)  = E_{2,\text{hyb}}^{\text{BH}} a(d+1,d+1), 
\label{HYBplus} \\
&&\left[ {\cal E}_2^{\text{BH}} + U  \right] a(d\!+\!L\!-\!1,d\!+\!L\!-\!1) - \sqrt{2} J a(d\!+\!L\!-\!1,d)  = E_{2,\text{hyb}}^{\text{BH}} a(d\!+\!L\!-\!1,d\!+\!L\!-\!1),
\label{HYBminus}
\end{eqnarray}
with $a(d,d)=a(d+L,d+L)=0$.
By manipulating the ansatz
\begin{eqnarray}
a(x,y) &=& \left[ C e^{i\theta (y+1)} + C' e^{-i\theta (y+1)} \right] \delta_{x,d} 
\label{ahyb}\\
&+& \left[ C e^{i\theta (d+1)} + C' e^{-i\theta (d+1)} \right] \delta_{x,d+1} \delta_{y,d+1} 
+ \left[ C e^{i\theta (d+L+1)} + C' e^{-i\theta (d+L+1)} \right] \delta_{x,d+L-1} \delta_{y,d+L-1},
\nonumber
\end{eqnarray}
with $d+1\leq y \leq d+L-1$, in the equations above, we derive the relation between the coefficients $C$ and $C'$, 
\begin{equation}
C'=- e^{2 i \theta d} C \frac{(U-g) e^{i\theta} +J }{(U-g) e^{-i\theta} +J }
\label{CC'hyb}
\end{equation}
and the energies
\begin{equation}
E_{2,\text{hyb}}^{\text{BH}}=  {\cal E}_2^{\text{BH}}+ g  - 2 J \cos \theta,
\label{Ehyb}
\end{equation}
with $\theta $ obtained from 
\[
f(\theta) = f(-\theta), \hspace{0.6 cm} f(\theta)=
e^{i \theta L} \left[U-g +J e^{i\theta} \right]^2.
\]
In the case of exact resonance, $g=U$, the eigenstates are approximately those given by Eqs.~(\ref{Psilocfree}) and (\ref{thetalocfree}), but now with the sum in $y$ from $d+1$ to $d+L+1$ and $k = 1, 2, \ldots L+1$.

We might expect the BP-states to mix with the states from the hybrid band above in second order of perturbation theory through the states $\phi(d+2,d+2)$ and $\phi(d+L-2,d+L-2)$.  The BP on site $d+2$ [and similarly for site $d+L-2$] would either hop onto site $d+1$ or split into sites $(d,d+2)$. However, the probability amplitudes for these transitions are almost equal in magnitude and opposite in sign, therefore canceling out. This can be seen from Eqs.~(\ref{ahyb}) and (\ref{CC'hyb}), which lead to
\begin{equation}
a(d+1,d+1)+a(d,d+2) = \frac{2iC\sin(2\theta) e^{i\theta d}}{(U-g)e^{-i\theta} +J}
\left[ U - g +2 J \cos \theta \right].
\end{equation}
The sum above approaches zero when $U \approx g - 2 J \cos \theta$ to  zeroth order in $g^{-1}$, as used to obtain Eq.~(\ref{Ehyb}). 
The absence of hybridization is therefore a result of destructive quantum interference and guarantees the stability of the BPs.  A similar anti-resonance occurs also for BPs in the XXZ model \cite{SantosDykman2003,DykmanSantos2003}.

One can think of the phenomenon as a Fano anti-resonance, where the narrow band of BPs in the presence of a defect goes through the broad band of loc-free states without scattering. This absence of scattering  can be equivalently understood from the following analysis of the motion of the bound pair on site $d+2$ (the case of the site $d+L-2$ is completely similar). The Schr\"odinger equation for this pair is
\begin{equation}
\label{eq:site_d+2}
\left[ {\cal E}_2^{\text{BH}} + U +\frac{2J^2}{U} -E^{\text{BH}} \right] a(d+2,d+2)   -\frac{2J^2}{U}a(d+3,d+3) - \sqrt{2} J a(d+1,d+2) = 0,
\end{equation}
where $E^{\text{BH}} \approx {\cal E}_2^{\text{BH}} + U$. 
Above, we have kept the term of hopping to site $d+3$, as in Eq.~(\ref{BPextended}), but not the hopping to $d+1$; instead, we have the amplitude for the intermediate virtual state $\phi(d+1,d+2)$.
Even though this last term is nonresonant, it is directly coupled to the following resonant terms,
\begin{equation}
\label{eq:d+1_d+2}
\left({\cal E}_2^{\text{BH}} -E^{\text{BH}}\right) a(d+1,d+2) - \sqrt{2} J \left[a(d+1,d+1) + a(d+2,d+2)\right]  - Ja(d,d+2)  = 0.
\end{equation}
Notice that $a(d+1,d+3)$ is nonresonant. 

We now substitute Eq.~(\ref{eq:d+1_d+2}) into Eq.~(\ref{eq:site_d+2}), taking into account that 
$ E_{}^{\text{BH}} - {\cal E}_2^{\text{BH}} \approx U$. We also use the Schr\"odinger equations for the states near the defect,  with energies $E_{}^{\text{BH}}$,
\begin{eqnarray}
&&\left[ {\cal E}_2^{\text{BH}} + g - E^{\text{BH}}\right] a(d,d+1) - \sqrt{2}J a(d+1,d+1) - Ja(d,d+2)] = 0, 
\label{HYBextended} \\
&&\left[ {\cal E}_2^{\text{BH}} + U -E^{\text{BH}} \right] a(d+1,d+1) - \sqrt{2} J a(d,d+1) = 0,
\label{HYBplus}
\end{eqnarray}
to rewrite $a(d,d+2)$ in terms of $a(d+1,d+1)$. [Eq.(27) and (36) coincide if one sets $E^{\text{BH}}=E^{\text{BH}}_{\text{2,hyb}}$]. This finally leads to
\begin{eqnarray}
\label{eq:pair_decoupling}
&&\left[ {\cal E}_2^{\text{BH}} + U +\frac{4J^2}{U} -  E_{}^{\text{BH}} \right ] a(d+2,d+2) -\frac{2J^2}{U}a(d+3,d+3)
\nonumber\\
&&\approx - U^{-1}\left[ {\cal E}_2^{\text{BH}} + U  -  E_{}^{\text{BH}} \right ]\left[ {\cal E}_2^{\text{BH}} + g  -  E_{}^{\text{BH}} \right ] a(d+1,d+1).
\end{eqnarray}
For propagating BPs, the energy $E_{}^{\text{BH}} $ lies within the band centered at $ {\cal E}_2^{\text{BH}} + U +4J^2/U$ with width $8J^2/U$ (see Fig.~\ref{fig:diagram}), therefore the coefficient at $a(d+1,d+1)$ in Eq.~(\ref{eq:pair_decoupling}) is $\sim (J^2/U)|g-U|/U\ll J^2/U$. This means that the propagating bound pairs do not hop onto site $d+1$ (nor onto site $d+L-1$). They are elastically reflected from the defect and do not hybridize with the loc-free states. 

\begin{figure}[htb]
\vskip 0.4cm
\includegraphics[width=4.0in]{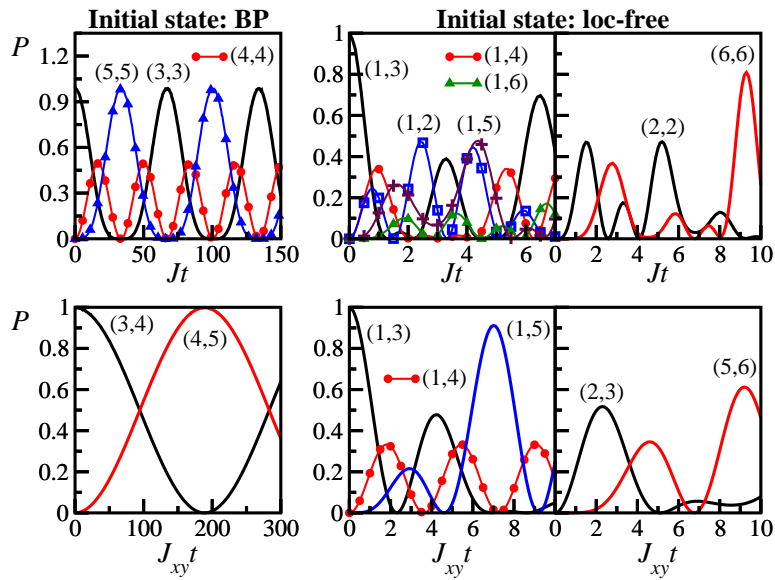}
\caption{Time evolution of two-particle excitations in the BH model (top) and XXZ model (bottom) for a chain of length $L=6$ with a defect on site $d=1$. The curves give the probability $P$ to find a state $\phi(x,y)$ at time $t$. The defect excess energy is exactly equal to the interaction strength: $g=U$ for the Bose-Hubbard model and $g=J_z$ for the XXZ model. The repulsive interaction is strong, $U/J=J_z/J_{xy}=30$, as in the experiments of Ref.~\cite{Winkler2006}. Left panels: the initial states are bound pairs (doublons) with wave functions $\phi(3,3)$ (top) and $\phi(3,4)$ (bottom). Middle and right panels: the initial state is a loc-free state with wave function $\phi(1,3)$ (top and bottom). The bound-pair states on the right panel are localized on sites next to the defect (``surface states"), no BPs in the bulk have an appreciable amplitude.}
\label{fig1}
\end{figure}

Figure~\ref{fig1} illustrates the decoupling of the resonating BP- and loc-free-bands for the BH model (top panels) and the XXZ model (bottom panels). A small chain is considered to facilitate the visualization of the states involved. The figure shows the time evolution of the probability $P$ to find a state $\phi(x,y)$ in the full two-excitation wave function of the system $\Psi(t)$, with $P\equiv P(x,y;t)=|\langle \phi(x,y)|\Psi(t)\rangle |^2$. The initial states on the left panels are BPs away from the defect site, whereas for the middle and right panels they are a loc-free-state. On the left, only BP states are seen, confirming the stability of the pair of bound excitations even when its splitting is not prohibited by energy conservation. The pairs move back and forth through the chain, being reflected by the defect. The middle and right panels show that the states that can display an appreciable coupling with the initial state are of two kinds: (i) the states with one excitation on the defect site and the other moving back and forth in the chain (middle panels), and (ii) the BPs localized next to the defect (right panels).

\section{Antiresonance stabilization of bound pairs in a defect-free system with many excitations}
\label{TwoBPs}

We now consider more than one BP in a defect-free chain, $g=0$. The goal is to analyze whether the pairs remain stable when colliding or can form larger clusters of bound excitations. 
We show that, similarly to the previous section, destructive quantum interference prevents BPs from merging into bound triples of excitations, even when this transition is energetically favorable, and therefore, as long as their density is not too high, BPs remain long-lived many-particle states. 

Of particular interest for not too high densities is the decay of BPs as a result of a collision of two BPs. In the BH-model with strong interaction, $U \gg J$, the energy of a state with two BPs is $\approx 2U$. Since the total number of excitations is conserved, the two BPs could decay into a bound triple of excitations and a freely propagating excitation or in a bound quartet of excitations. The energy of a triple of excitations on the same site is $\approx 3U$ and the energy of a quartet is $\approx 6U$. Therefore, by energy conservation, a collision of two BPs (two doublons) cannot lead to their decay in this model. Decay processes would require a collision of at least 3 doublons. 
In the XXZ model, on the other hand, the decay of BPs could indeed result from a collision of two BPs. In this case, the energies of the bound triple (3 upward spins on neighboring sites) and of two BPs are $\approx 2J_z$.  In principle, the two BPs could then combine into a bound triple and a free excitation. The amplitde for this transition is, however, very small, as we show below.

To study the stability of the BPs in an ideal XXZ chain, we present a four-excitation wave function as 
\begin{equation}
\label{eq:psi_4}
\psi_4^{\text{XXZ}}=\sum\nolimits_{x_1<x_2<x_3<x_4}a(x_1,x_2,x_3,x_4)\phi(x_1,x_2,x_3,x_4),
\end{equation}
where $\phi(x_1,x_2,x_3,x_4)$ is the state of the system with spins on sites $x_1,\ldots,x_4$ pointing up, whereas all other spins are pointing down. An eigenstate with two spatially separated bound pairs is described by the wave function (\ref{eq:psi_4}) where the coefficients have the structure $a(x,x+1,y,y+1)$ with $y \geq x+3$. These states form a narrow band centered at $\approx {\cal E}_4^{\text{XXZ}}+2J_z$ with typical width $\sim 2J^2_{xy}/J_z$, cf. Table~II.

A bound triple and an unbound free excitation are described by the wave function (\ref{eq:psi_4}) with coefficients that have the structure $a(x,x+1,x+2,y)$ with $y\geq x+4$. They also form a band centered at  $\approx {\cal E}_4^{\text{XXZ}}+2J_z $, but with a much larger width $\sim J_{xy}$, which is determined primarily by the hopping of the free excitation.

Two BPs separated by only one site are directly coupled to a triple. To the leading order in $J_{xy}/J_z$, the Schr\"odinger equation for the corresponding resonating amplitudes for energies $E^{\text{XXZ}}$ reads
\begin{equation}
\label{eq:triple_mixing}
\left({\cal E}_4^{\text{XXZ}}+2J_z-E^{\text{XXZ}}\right)a(x,x+1,x+2,x+4)-\frac{J_{xy}}{2} [a(x,x+1,x+3,x+4)+a(x,x+1,x+2,x+5)]=0.
\end{equation}
Here, the triple is on sites $(x,x+1,x+2)$ and the BPs are on sites $(x,x+1)$ and $(x+3,x+4)$.

BPs separated by more than one site could mix with triples and with BPs separated by one site in second order of perturbation theory, but the amplitudes of these processes turn out to be very small. To describe what happens, we consider the BPs on sites $(x,x+1, x+4,x+5)$ and show that they have a very small amplitude of hopping onto sites $(x,x+1,x+3,x+4)$ [or equivalently onto site $(x+1,x+2,x+4,x+5)$], that is, into the state strongly admixed with bound triples. To this end we write the Schr\"odinger equation
\begin{eqnarray}
\label{eq:BPs_next_neighb}
&&\left({\cal E}_4^{\text{XXZ}}+2J_z + \frac{J^2_{xy}}{2J_z}-E^{\text{XXZ}}\right)a(x,x+1,x+4,x+5)
+\frac{J^2_{xy}}{4J_z} \big[ a(x,x+1,x+5,x+6) + a(x-1,x,x+4,x+5)\big] \nonumber\\
&&  - \frac{1}{2}J_{xy}  \big[ a(x,x+1,x+3,x+5)+ a(x,x+2,x+4,x+5) \big] =0.
\end{eqnarray}
The last two terms in this equation are nonresonant and correspond to a virtual dissociation of one of the BPs. They describe the intermediate states that lead to hopping of the BPs toward each other. The Schr\"odinger equation for $a(x,x+1,x+3,x+5)$, taking into account Eq.~(\ref{eq:triple_mixing}) and the fact that $E^{\text{XXZ}}\approx {\cal E}_4^{\text{XXZ}}+2J_z$ for two-BP states, leads to
\begin{eqnarray}
\label{eq:BPs_internal_hopping}
\frac{1}{2}J_{xy}a(x,x+1,x+3,x+5)\approx -\frac{J_{xy}}{2J_z}\left({\cal E}_4^{\text{XXZ}}+2J_z-E^{\text{XXZ}}\right)a(x,x+1,x+2,x+4).
\end{eqnarray}
Function $a(x,x+2,x+4,x+5)$ is similarly expressed in terms of the bound-triple amplitude $a(x+1,x+3,x+4,x+5)$.
 
It is immediately seen from Eqs.~(\ref{eq:BPs_next_neighb}) and (\ref{eq:BPs_internal_hopping}) that, for the characteristic energies $E^{\text{XXZ}}$ lying within the bandwidth of the BP excitations, $|E^{\text{XXZ}}- ({\cal E}_4^{\text{XXZ}}+2J_z)| \lesssim J_{xy}^2/J_z$, the admixture of the states with BPs separated by two sites and the states with bound triples is parametrically smaller than the bandwidth of two-BP-states. Physically, this means that once two BPs approach each other, they are reflected before they become closer than two empty lattice sites. Such reflection is due to a destructive quantum interference between the bound triple and two-BP states separated by one site and can also be interpreted in terms of Fano anti-resonance.

\begin{figure}[htb]
\vskip 0.2cm
\includegraphics[width=4.0in]{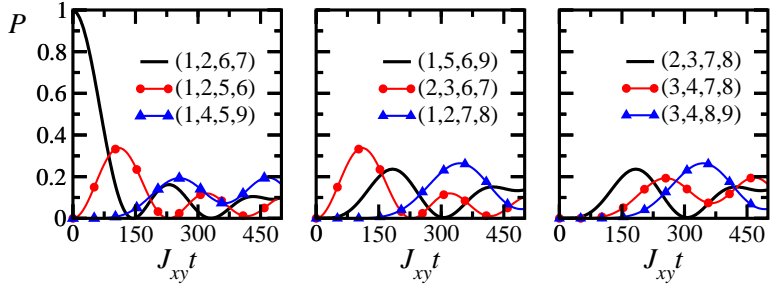}
\caption{Time evolution of 4-excitation states in a defect-free XXZ chain of length  $L=9$ for large repulsive interaction $J_z/J_{xy}=30$. The curves show the probabilities $P = |\langle\phi(x_1,x_2,x_3,x_4)|\psi_4^{\text{XXZ}}(t)\rangle|^2$ for different sites. The initial state $\phi(1,2,6,7)$ couples only with states from the two-BP-band, so only states with BPs separated by two or more sites are seen.}
\label{fig2}
\end{figure}

The time evolution of the four-excitation states in the XXZ model with a large repulsive interaction is illustrated in Figure~\ref{fig2} for a defect-free chain with nine sites and periodic boundary conditions. In this system, there are nine states with two BPs separated by two or more sites. The initial state considered has the BPs separated by three sites, $\phi(1,2,6,7)$. These BPs move along the chain and reflect before the inter-pair distance becomes equal to 1. In the figure, only the nine curves corresponding to the states of the two-BP-band participate in the evolution, confirming the decoupling between this band and the hybrid band of triples and BPs separated by one site. Not a single state with three excitations next to each other or with two BPs separated by one site had an appreciable amplitude in the simulation for the studied time range.  This is in full agreement with the analytical results.

\section{Conclusion}

We studied the dynamics of bound pairs of excitations in one-dimensional systems with strong repulsive short-range interaction described by the Bose-Hubbard and the anisotropic Heisenberg models. In the first system, the bound pair is formed by two atoms occupying the same site. Such state has become known as doublon. In the second system the bound pair corresponds to parallel spins placed on neighboring sites. For strong interaction, the bandwidth of bound pairs in an ideal chain is much smaller than the bandwidth of states with free excitations, so the bands of unbound and bound excitations are well separated.

Of primary interest was the question of stability of the bound pairs in two possible decay scenarios. One refers to a chain with one bound pair and a single on-site defect. The energy of the defect can be tuned close to the interaction energy of the bound pair so as to allow energy conservation in the scattering of the BP off the defect.  The second scenario refers to an ideal chain where the total energy of two BPs is equal to the sum of the energy of a bound triple of excitations and the energy of a propagating single excitation, so that we might expect the BPs to merge into a triple.

Our central result is the observation that, for strong interaction, both processes are suppressed. This is because there are two intermediate states that can mediate the decay process of the pairs and these states have essentially equal  and opposite in sign amplitudes. Such destructive quantum interference leads to the long-time stability of the pairs of bound excitations. 

The stability of bound pairs of excitations with strong negative binding energy has immediate bearing on the possibility of observing them in the experiment. In experiments with cold atoms it is possible both to create a ``defect" in the optical lattice and to control its energy. It is also possible to change the density of bound pairs. An observation of the stability of the pairs would be a direct demonstration of quantum antiresonance in the system.

\begin{acknowledgments}
LFS was supported by the NSF under grants DMR-1147430 and DMS-1107444. MID was supported by the NSF through grant EMT/QIS 082985
\end{acknowledgments}


\begin{thebibliography}{32}
\bibitem{footnote} The sign of $J_{xy}$ does not affect the results that we discuss; we chose it here opposite to that in \cite{SantosDykman2003,DykmanSantos2003}.
\end{thebibliography}
\end{document}